\documentclass[reprint,prd,aps,showpacs, showkeys,superscriptaddress]{revtex4-1}
\usepackage{amsmath}
\usepackage{graphicx}
\usepackage{hyperref}
\usepackage{epstopdf}
\usepackage{multirow}
\usepackage{subfigure}
\usepackage{color}
\usepackage{ulem}
\usepackage{gensymb}

\usepackage{color}
\usepackage{amssymb}
\definecolor{darkblue}{rgb}{0,0,0.5}

\usepackage{hyperref}  
\hypersetup{
    colorlinks=true, 
    linkcolor=darkblue,  
    urlcolor=darkblue,
    citecolor=darkblue,
    filecolor=darkblue,
}

\def\nuebar{\overline{\nu}_e}

\def\frac#1#2{\textstyle{{{#1} \over {#2}}}}

\def\lsim{\mathrel{\rlap{\lower4pt\hbox{\hskip1pt$\sim$}}
    \raise1pt\hbox{$<$}}}
\def\gsim{\mathrel{\rlap{\lower4pt\hbox{\hskip1pt$\sim$}}
    \raise1pt\hbox{$>$}}}
\def\Re{\hbox{Re}\,}
\def\Im{\hbox{Im}\,}

\newcommand{\beq}{\begin{equation}}
\newcommand{\eeq}{\end{equation}}
\newcommand{\bea}{\begin{eqnarray}}
\newcommand{\eea}{\end{eqnarray}}
\newcommand{\bse}{\begin{subequations}}
\newcommand{\ese}{\end{subequations}}

\def\to{\rightarrow}

\def\As#1{({\cal A}_s)_{#1}}
\def\Ac#1{({\cal A}_c)_{#1}}
\def\Bs#1{({\cal B}_s)_{#1}}
\def\Bc#1{({\cal B}_c)_{#1}}
\def\C#1{({\cal C})_{#1}}

\def\indx{{\bar{c}\bar{d}}}
\def\idxee{\bar{e}\bar{e}}
\def\idxmm{\bar{\mu}\bar{\mu}}
\def\idxtt{\bar{\tau}\bar{\tau}}
\def\idxem{\bar{e}\bar{\mu}}
\def\idxet{\bar{e}\bar{\tau}}
\def\idxmt{\bar{\mu}\bar{\tau}}

\begin{document}

\title{Search for a time-varying electron antineutrino signal at Daya Bay}

\newcommand{\ECUST}{\affiliation{Institute of Modern Physics, East China University of Science and Technology, Shanghai}}
\newcommand{\IHEP}{\affiliation{Institute~of~High~Energy~Physics, Beijing}}
\newcommand{\Wisconsin}{\affiliation{University~of~Wisconsin, Madison, Wisconsin 53706}}
\newcommand{\Yale}{\affiliation{Wright~Laboratory and Department~of~Physics, Yale~University, New~Haven, Connecticut 06520}} 
\newcommand{\BNL}{\affiliation{Brookhaven~National~Laboratory, Upton, New York 11973}}
\newcommand{\NTU}{\affiliation{Department of Physics, National~Taiwan~University, Taipei}}
\newcommand{\NUU}{\affiliation{National~United~University, Miao-Li}}
\newcommand{\Dubna}{\affiliation{Joint~Institute~for~Nuclear~Research, Dubna, Moscow~Region}}
\newcommand{\CalTech}{\affiliation{California~Institute~of~Technology, Pasadena, California 91125}}
\newcommand{\CUHK}{\affiliation{Chinese~University~of~Hong~Kong, Hong~Kong}}
\newcommand{\NCTU}{\affiliation{Institute~of~Physics, National~Chiao-Tung~University, Hsinchu}}
\newcommand{\NJU}{\affiliation{Nanjing~University, Nanjing}}
\newcommand{\TsingHua}{\affiliation{Department~of~Engineering~Physics, Tsinghua~University, Beijing}}
\newcommand{\SZU}{\affiliation{Shenzhen~University, Shenzhen}}
\newcommand{\NCEPU}{\affiliation{North~China~Electric~Power~University, Beijing}}
\newcommand{\Siena}{\affiliation{Siena~College, Loudonville, New York  12211}}
\newcommand{\IIT}{\affiliation{Department of Physics, Illinois~Institute~of~Technology, Chicago, Illinois  60616}}
\newcommand{\LBNL}{\affiliation{Lawrence~Berkeley~National~Laboratory, Berkeley, California 94720}}
\newcommand{\UIUC}{\affiliation{Department of Physics, University~of~Illinois~at~Urbana-Champaign, Urbana, Illinois 61801}}
\newcommand{\SJTU}{\affiliation{Department of Physics and Astronomy, Shanghai Jiao Tong University, Shanghai Laboratory for Particle Physics and Cosmology, Shanghai}}
\newcommand{\BNU}{\affiliation{Beijing~Normal~University, Beijing}}
\newcommand{\WM}{\affiliation{College~of~William~and~Mary, Williamsburg, Virginia  23187}}
\newcommand{\Princeton}{\affiliation{Joseph Henry Laboratories, Princeton~University, Princeton, New~Jersey 08544}}
\newcommand{\VirginiaTech}{\affiliation{Center for Neutrino Physics, Virginia~Tech, Blacksburg, Virginia  24061}}
\newcommand{\CIAE}{\affiliation{China~Institute~of~Atomic~Energy, Beijing}}
\newcommand{\SDU}{\affiliation{Shandong~University, Jinan}}
\newcommand{\NanKai}{\affiliation{School of Physics, Nankai~University, Tianjin}}
\newcommand{\UC}{\affiliation{Department of Physics, University~of~Cincinnati, Cincinnati, Ohio 45221}}
\newcommand{\DGUT}{\affiliation{Dongguan~University~of~Technology, Dongguan}}
\newcommand{\XJTU}{\affiliation{Department of Nuclear Science and Technology, School of Energy and Power Engineering, Xi'an Jiaotong University, Xi'an}}
\newcommand{\UCB}{\affiliation{Department of Physics, University~of~California, Berkeley, California  94720}}
\newcommand{\HKU}{\affiliation{Department of Physics, The~University~of~Hong~Kong, Pokfulam, Hong~Kong}}
\newcommand{\UH}{\affiliation{Department of Physics, University~of~Houston, Houston, Texas  77204}}
\newcommand{\Charles}{\affiliation{Charles~University, Faculty~of~Mathematics~and~Physics, Prague}} 
\newcommand{\USTC}{\affiliation{University~of~Science~and~Technology~of~China, Hefei}}
\newcommand{\TempleUniversity}{\affiliation{Department~of~Physics, College~of~Science~and~Technology, Temple~University, Philadelphia, Pennsylvania  19122}}
\newcommand{\CUC}{\affiliation{Instituto de F\'isica, Pontificia Universidad Cat\'olica de Chile, Santiago}} 
\newcommand{\CGNPG}{\affiliation{China General Nuclear Power Group, Shenzhen}}
\newcommand{\NUDT}{\affiliation{College of Electronic Science and Engineering, National University of Defense Technology, Changsha}} 
\newcommand{\IowaState}{\affiliation{Iowa~State~University, Ames, Iowa  50011}}
\newcommand{\ZSU}{\affiliation{Sun Yat-Sen (Zhongshan) University, Guangzhou}}
\newcommand{\CQU}{\affiliation{Chongqing University, Chongqing}} 
\newcommand{\BCC}{\altaffiliation[Now at ]{Department of Chemistry and Chemical Technology, Bronx Community College, Bronx, New York  10453}} 
\author{D.~Adey}\IHEP
\author{F.~P.~An}\ECUST
\author{A.~B.~Balantekin}\Wisconsin
\author{H.~R.~Band}\Yale
\author{M.~Bishai}\BNL
\author{S.~Blyth}\NTU\NUU
\author{D.~Cao}\NJU
\author{G.~F.~Cao}\IHEP
\author{J.~Cao}\IHEP
\author{J.~F.~Chang}\IHEP
% ...

\collaboration{The Daya Bay Collaboration}\noaffiliation
\date{\today}

\begin{abstract}
A search for a time-varying $\bar{\nu}_{e}$ signal was performed with 621 days of data acquired by the Daya Bay Reactor Neutrino Experiment over 704 calendar days. The time spectrum of the measured $\nuebar$ flux normalized to its prediction was analyzed with a Lomb-Scargle periodogram, which yielded no significant signal for periods ranging from 2 hours to nearly 2 years.   
The normalized time spectrum was also fit for a sidereal modulation under the Standard Model extension (SME) framework to search for Lorentz and CPT violation (LV-CPTV). Limits were obtained for all six flavor pairs $\idxem,\idxet$, $\idxmt$, $\idxee,\idxmm$ and $\idxtt$ by fitting them one at a time, constituting the first experimental constraints on the latter three. Daya Bay's high statistics and unique layout of multiple directions from three pairs of reactors to three experimental halls allowed 
the simultaneous constraint of individual SME LV-CPTV coefficients without assuming others contribute negligibly, a first for a neutrino experiment. 

\end{abstract}


\maketitle

\section{Introduction}
Some scenarios of physics beyond the Standard Model (SM) predict a time-varying probability of neutrino oscillation. Among these are models in which ultralight scalar dark matter couples to neutrinos, inducing periodic variations in the mass splittings and mixing angles~\cite{Scenario1, Scenario3}. Other models involve Lorentz symmetry violation (LV), which is suggested as a signature of Planck scale phenomenology~\cite{LV1,LV2,CPTV,AIPProceeding} and which could be accompanied by CPT violation (CPTV)~\cite{LV2, Greenberg}. 

The Standard Model extension (SME)~\cite{PRD55_6760, PRD69_105009, LecN} was introduced as an effective theory that maintains the usual gauge structure and properties of the SM such as renormalizability, but adds all the possible terms constructed with SM fields that introduce Lorentz symmetry breaking. By predicting a set of testable signatures in various areas of physics, it provides a connection between experimental research and more fundamental theories extending to the Planck scale.

In the neutrino sector, the violation of rotation symmetry in the SME causes deviations from standard oscillation probabilities derived from the Pontecorvo-Maki-Nakagawa-Sakata (PMNS) matrix~\cite{PMNS} that depend on propagation direction. This would produce a time-varying neutrino oscillation probability associated with the Earth's orbital and rotational movement relative to the fixed stars, and therefore a period of a sidereal day (23~h 56~min 4.09~sec). Accordingly, a sidereal time dependence has been sought in the oscillation probability of
accelerator neutrinos~\cite{LSNDPRD72_076004,MiniBooNEPLB718,T2K_111101,MINOSPRL101_151601,MINOSPRL105_151601,MINOSPRD85_031101}, atmospheric neutrinos~\cite{ICEcubePRD82_112003} and reactor neutrinos~\cite{DoubleChoozPRD86_112009}. The SME also predicts  deviations from the standard $L/E$ oscillation behavior. 
The oscillated neutrino energy spectrum of atmospheric neutrinos has been examined for such a distortion both in the Super Kamiokande~\cite{SuperK_Energy} and IceCube~\cite{IceCube_Energy} experiments. No positive LV or CPTV signal has yet been observed, and neutrino oscillation experiments have set some of the most stringent limits on the violation of these fundamental symmetries of nature, down to the level of $10^{-28}$~\cite{IceCube_Energy}. 

The Daya Bay reactor neutrino experiment~\cite{DYBNIM2016} has recently produced the most precise measurements of reactor electron antineutrino ($\overline{\nu}_e$) disappearance at short baselines~\cite{DYBPRL2012,DYBCPC2013,DYB_nH_PRD2014,DYBPRL2015,DYB_nH_PRD2016,DYBPRD2017}. In neutrino oscillation experiments, time-dependent LV-CPTV effects are amplified with distance, and Daya Bay's baselines ($<2$~km) are relatively short compared to many other neutrino oscillation experiments. However, Daya Bay has accumulated the largest sample of reactor $\overline{\nu}_e$'s to date. Moreover, it has a unique experimental layout comprising different well-known neutrino propagation directions. Both of these factors make it an excellent experiment to search for a time-varying $\bar{\nu}_e$ signal and LV-CPTV effects. 

This paper first describes a generic search for an unpredicted periodicity in the $\bar{\nu}_e$ rates measured at Daya Bay using the Lomb-Scargle method~\cite{LSmethod}. This analysis, which yields no positive results, has the potential to identify the presence of an unexpected time-variant source of $\bar{\nu}_e$'s.  The paper then presents a targeted search for a sidereal time modulation in the context of the SME, producing limits on the individual coefficients that characterize the theory.

\section{Antineutrino Data Set}

\subsection{Experiment description}
\begin{figure}[!htbp]
 \begin{center}
    \includegraphics[width=\columnwidth]{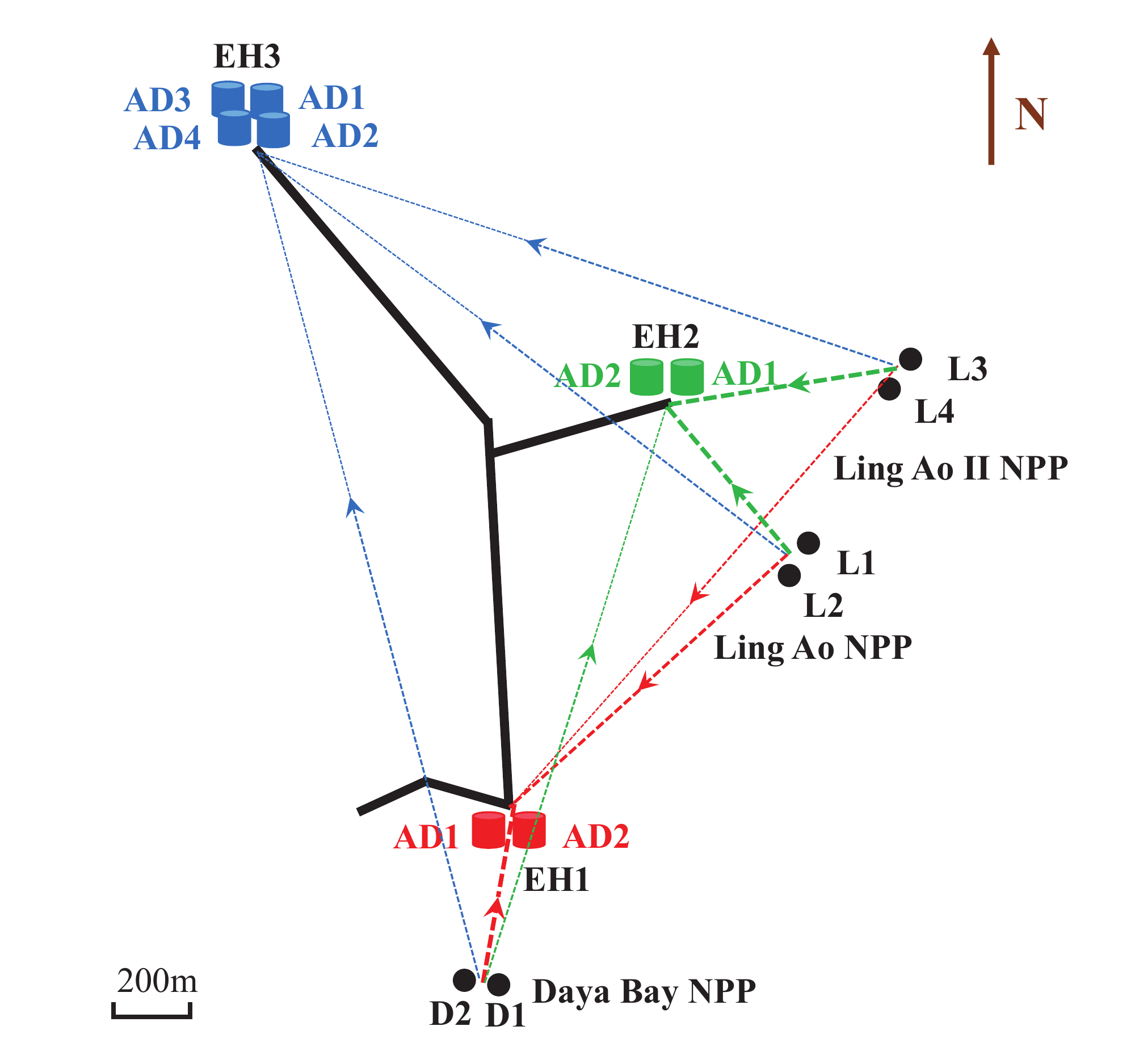}
    \caption{Layout of Daya Bay reactor cores (black dots) and antineutrino detectors (colored cylinders). The six reactor cores are located in three nuclear power plants (NPPs). The dashed lines and arrows show the multiple $\overline{\nu}_e$ `beams' from the different reactors to the three experimental halls (EHs). The solid black lines represent the underground tunnels.}
\label{Fig:Map}
\end{center}
\end{figure}

\begin{figure*}[!htbp]
\begin{center}
     \includegraphics[width=0.99\textwidth]{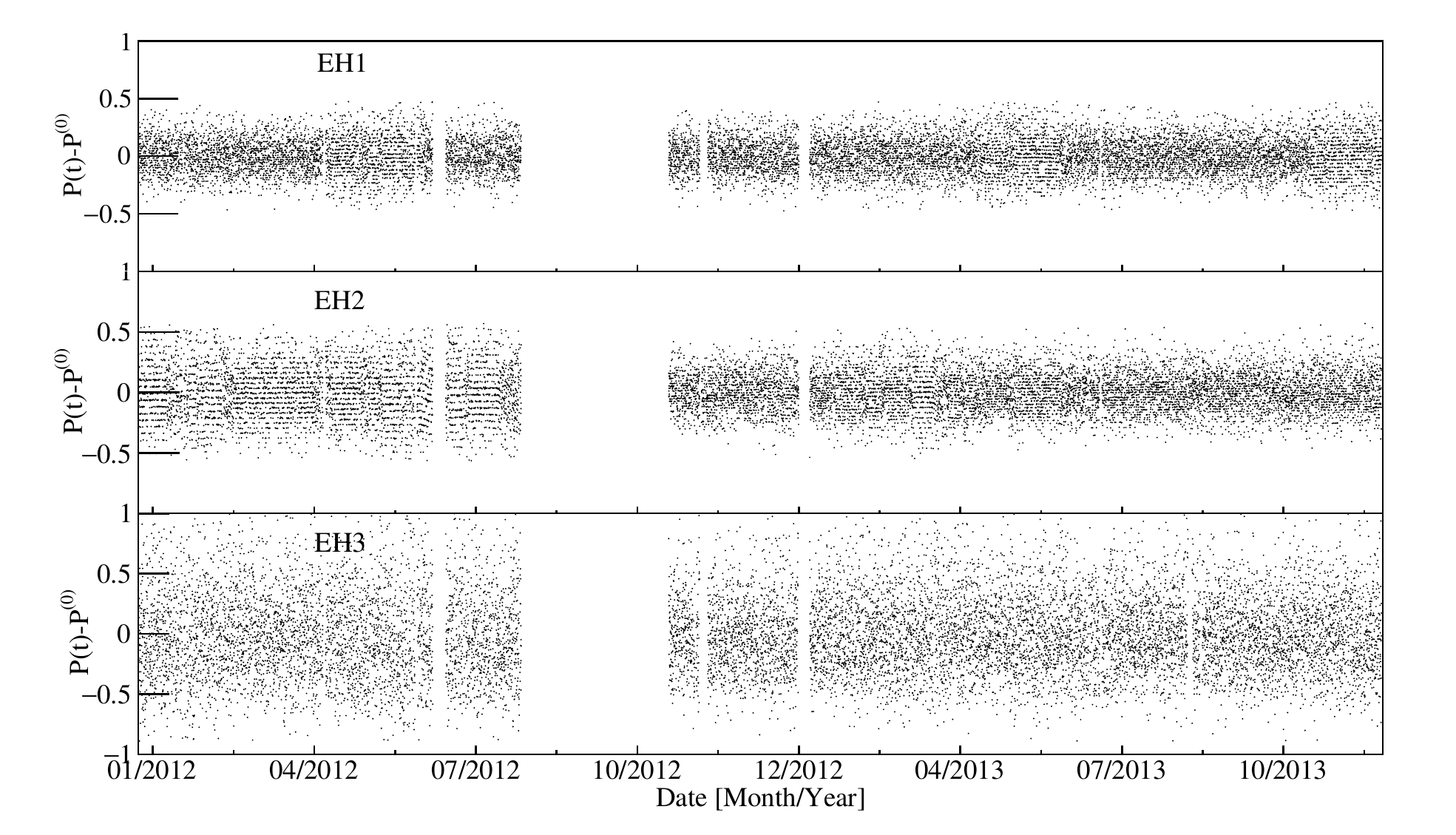}
    \caption{Measured residual survival probability as a function of real time in sidereal hour bins for each of the experimental halls over 704 solar days of data acquisition. No error bars are plotted to avoid cluttering. The gaps correspond to breaks in physics data acquisition, the largest of which occurred in 2012 between July 28 and October 19, due to the installation of two additional ADs. Discrete steps along the vertical axis, which are most apparent in the first 7 months of EH2, are due to the low statistics acquired from a single detector in 1 sidereal hour, while the data from most other periods were averaged among multiple detectors. 
    }
\label{Fig:Data_hourly}
\end{center}
\end{figure*}
The Daya Bay reactor complex consists of three nuclear power plants (Daya Bay, Ling Ao, Ling Ao II), each with two reactors.  The emitted $\overline{\nu}_e$ flux is sampled in eight identically designed antineutrino detectors (ADs) located in three experimental halls (EHs), as shown in 
Fig.~\ref{Fig:Map}.
Each AD is filled with 20 tons of gadolinium-doped liquid scintillator enclosed by 22 tons of undoped liquid scintillator and 40 tons of mineral oil.
Scintillation light is detected by 192 photomultiplier tubes (PMTs). The ADs are immersed in pure water pools that are instrumented with PMTs, providing shielding and serving as cosmogenic muon detectors. 
Candidate events independently trigger each detector and are read out by custom front-end electronics. For the ADs, a readout window of $1.2~\mu s$ of data is initiated when the number of PMTs with an above-threshold signal is greater than 45 or the synchronous analog sum of charge output by the PMTs is larger than a value corresponding to $\sim$$0.4$~MeV~\cite{DYBNIM2016}. 
The clock for the readout electronics and trigger systems runs at 40 MHz and is synchronized to a global 10 MHz signal
generated by a rubidium oscillator further synchronized to absolute coordinated universal time (UTC) with a global positioning system (GPS) receiver. 
Further information about the Daya Bay experiment can be found in Ref.~\cite{DYBNIM2016}.

\subsection{Antineutrino signal and backgrounds}

\begin{figure*}[!htbp]

     \includegraphics[width=0.99\textwidth]{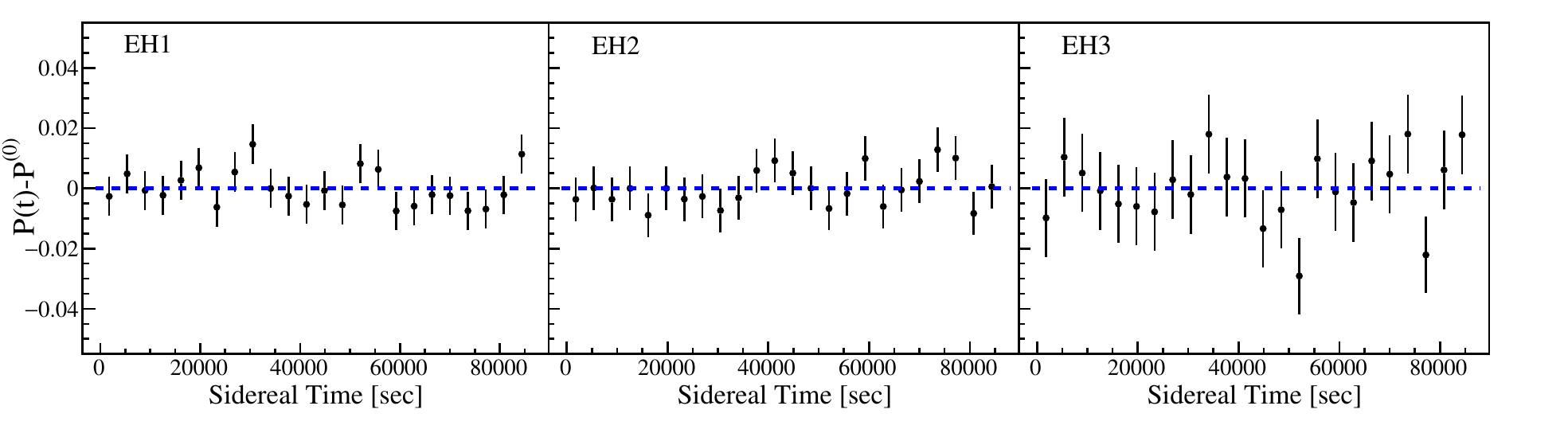}
    \caption{Measured residual survival probability in 24 sidereal hour bins spanning 704 solar days. Statistical and systematic uncertainties are considered for each bin.}
\label{Fig:Data_24}
\end{figure*}

The data set used in this study corresponds to a total exposure of 621 days distributed over 704 solar days (705.5 sidereal days), from December 24, 2011 to November 27, 2013. Data taking began with 6 ADs and continued for 217 days, pausing from July 28, 2012 to October 19, 2012 for the installation of the final two ADs, one in EH2 and the other in EH3.
Each physics run lasted as long as 72 hours. 
A 3-hour interruption of normal data acquisition occurred almost every Friday to calibrate the detectors. 

Electron antineutrinos were detected via the inverse beta decay (IBD) reaction, $\nuebar +p \to e^{+} + n$,  where the energy loss of the $e^{+}$ in the scintillator and its subsequent annihilation provided a prompt scintillation light signal followed by a delayed light signal from the neutron capture on gadolinium. IBD candidates were selected by requiring prompt-delayed pairs to have specific energies ($0.7<E_{\mathrm{prompt}}<12.0$~MeV, $6.0<E_{\mathrm{delayed}}<12.0$~MeV) and time separation ($1<\Delta t < 200$~$\mathrm{\mu s}$). Selected events were also required not to have been preceded by a muon candidate. A multiplicity cut was applied to ensure that only isolated prompt-delayed pairs were selected. Two slightly different IBD selections based on these criteria were used in two independent analyses, which also estimated backgrounds differently. 
Distinct muon veto and multiplicity cut efficiencies were accurately assessed from muon and random background rates as a function of time, and applied in the estimation of the IBD rates. The time dependence of these efficiencies was negligible. 

The total background amounted to less than $3\%$ of the total IBD candidate samples and was dominated by accidental coincidences. Both analyses precisely determined this background hourly from the measured rates of uncorrelated signals and subtracted it from the IBD samples. One analysis also considered the small variations in the fast neutron and $^{9}$Li$/^{8}$He correlated backgrounds, which were determined for the full period and then estimated hourly by scaling them with the measured muon rate. The same was done for the background caused by neutrons from the $^{241}$Am-$^{13}$C calibration sources, but scaling with the hourly single neutron rate. The slight time dependence in these backgrounds was found to contribute $<0.01\%$ of the total variation in IBD rates, and thus had a negligible impact on the results presented in this paper.  

The $\overline{\nu}_e$ oscillation probability was determined as the ratio of the measured IBD rate to the predicted IBD rate assuming no oscillation.
The measured IBD rate in each hall was determined hourly by dividing the number of background-subtracted IBD events by the data acquisition livetime, correcting for the loss of time caused by the muon veto and multiplicity cuts. This rate was then divided by the hourly expectation determined as in Ref.~\cite{DYBCPC2017} but with a livetime-weighted linear interpolation of daily thermal power data, yielding the measured survival probability $P(t)$. The time average of $P(t)$ was normalized to the Lorentz-invariant three-neutrino survival probability $P^{(0)}$ measured by Daya Bay in Ref.~\cite{DYBPRL2015} with the same data set, and $P^{(0)}$ was subtracted to give the residual survival probability $R(t) \equiv P(t)-P^{(0)}$. 
This quantity is shown in Figs.~\ref{Fig:Data_hourly} and \ref{Fig:Data_24}, the former vs. real time in hourly bins, and the latter accumulated in 24 sidereal hour bins~\cite{supmat}. The origin for the sidereal time was set to local midnight on the 2018 vernal equinox, a convention typically used by experiments searching for LV and CPTV in the sun-centered frame. An integer number of sidereal days was subtracted to obtain the closest foregoing time to the start of data taking, yielding 2011/12/23 22:13:13.80705 UTC. This choice had no impact on the limits reported in Secs.~\ref{sec:ls} and~\ref{sec:smefit}.

\subsection{Uncertainties}\label{sec:unc}

Given the nature of this search, only uncertainties of quantities that varied over time were taken into account. These included statistical, reactor-related, and event selection uncertainties. The statistical uncertainty dominated the uncertainty in each EH, contributing at the level of 0.63\%, 0.71\%, and 1.26\% of $P(t)$ in each of the 24 time bins of Fig.~\ref{Fig:Data_24} for EH1, EH2 and EH3, respectively. 

The efficiency uncertainty was dominantly due to the delayed energy ($E_{\mathrm{delayed}}$) cut, and was inferred using the estimated stability of the energy scale. The energy scale was calibrated during data collection using spallation neutrons, and was found to vary within 0.2\% in all ADs~\cite{DYBNIM2016}. Variations in the number of target protons, amount of neutrons produced by IBD interactions outside the target that diffused into the target, and neutron capture time were estimated with the relationship between the density $\rho$ and temperature $T$ of the gadolinium-doped liquid scintillator: $\Delta \rho = -9.05\times 10^{-4} \Delta T$~\cite{LSDensityTemp}. 
The expected $0.045$\% change in $\rho$ based on the observed $0.5\degree$C variation in temperature was propagated to the uncertainties of these parameters. 
Uncertainties of all other selection efficiencies were less significant and conservatively inherited from the oscillation analysis~\cite{DYBPRL2015}. The overall uncertainty of the selection efficiency in the three halls was estimated to be $0.09\%$ of the survival probability $P(t)$ for each bin of Fig.~\ref{Fig:Data_24}. When combining the data of individual ADs in the same hall, correlations were considered. 

All the uncorrelated reactor-related uncertainties involved in the flux prediction, which included power, energy/fission, fission fraction, spent fuel and nonequilibrium corrections, totaled to $0.9\%$ and were conservatively treated as time dependent on a daily basis. The relative size of the reactor systematic with respect to the survival probability $P(t)$ for each bin in Fig.~\ref{Fig:Data_24} was 0.10\%, 0.09\%, and 0.08\% for EH1, EH2, and EH3, respectively. Correlations between the predicted fluxes at the three halls had negligible impact to the analyses presented. 

As discussed previously, background variation with time was found to be negligible.

\section{Analysis on Periodic Amplitudes}\label{sec:ls}

\begin{figure*}[]
     \includegraphics[ width=0.97\textwidth]{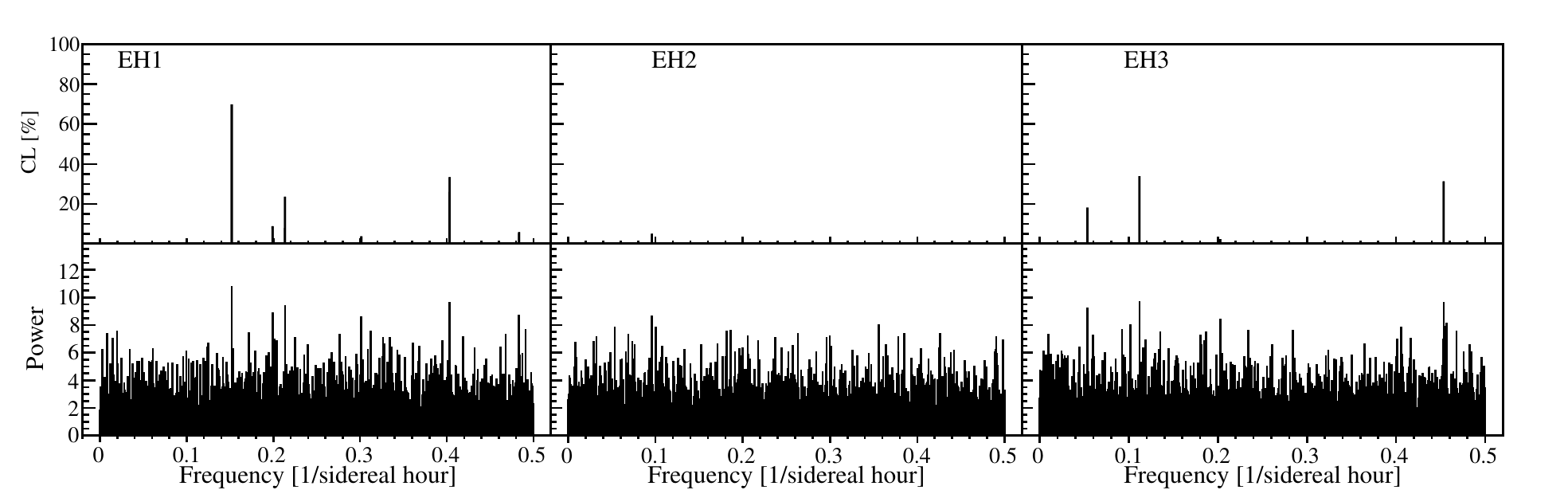}
    \caption{Lomb-Scargle powers (bottom) and confidence levels (top) for each experimental hall.}
\label{Fig:LS_CL}
\end{figure*}

A general search for a periodic signal within the measured residual survival probability was performed for each of the three experimental halls using the Lomb-Scargle (LS) periodogram~\cite{LSmethod}, which is a widely used technique for detecting periodic signals in unevenly sampled data. A periodogram was derived for each panel in Fig.~\ref{Fig:Data_hourly}, spanning a frequency range from 5.9$\times10^{-5}$ sidereal hour$^{-1}$ to  0.5 sidereal hour$^{-1}$. The normalized LS power for a frequency $f$ derived from $N$ data points $X_{j}$ at specific times $t_{j}$ can be estimated as~\cite{LSmethod}
\begin{equation}
\begin{split}
L(f) = &\dfrac{1}{2\sigma^{2}} \Biggl\{ \frac{[\sum^{N}_{j=1}(X_{j}-\overline{X}) \mbox{cos}(2\pi f (t_{j}-\tau)) ]^{2} }{\sum^{N}_{j=1} \mbox{cos}^{2}(2\pi f (t_{j}-\tau))}   \\
 &  + \frac{[\sum^{N}_{j=1}(X_{j}-\overline{X}) \mbox{sin}(2\pi f (t_{j}-\tau)) ]^{2} }{\sum^{N}_{j=1} \mbox{sin}^{2}(2\pi f (t_{j}-\tau))}   \Biggl\},
\end{split}
\end{equation}
with $\overline{X} \equiv \sum^{N}_{j=1} X_{j}/N$ and $\tau$ defined by $\mbox{tan}(4\pi f \tau) = \sum^{N}_{j=1} \mbox{sin}(4\pi f t_{j}) /  \sum^{N}_{j=1} \mbox{cos}(4 \pi f t_{j})$. The normalization is accomplished by dividing by the total variance, $\sigma^{2} \equiv \sum^{N}_{j=1}(X_{j}-\overline{X})^{2}/(N-1)$. The obtained values of $\sigma^2$ were $0.023$, $0.032$ and $0.112$ for EH1, EH2 and EH3, respectively. The bottom panels of Fig.~\ref{Fig:LS_CL} show the resulting LS powers for each frequency in each hall. 

It was noted in Ref.~\cite{LSnorm} that if the signal $X_{j}$ is purely white noise, then $L(f)$ follows an exponential probability distribution when normalized with $\sigma^2$. Accordingly, the significance of a given LS power can be determined with a confidence level (CL) defined as $(1-e^{-L(f)})^{M}$, where $M$ is the number of independent frequencies that are scanned. 
$M$ is nearly equal to the number of data points $M\approx N$ in the case of even sampling, but is {\it a priori} unknown for unevenly distributed samples. To estimate this number, 10,000 Monte Carlo data sets with statistical fluctuations were analyzed. The highest LS power $z$ in each data set was selected to construct a probability density function for each hall, which was then fit as $P(z)=M(1-e^{-z})^{M-1}e^{-z}$~\cite{LSnorm,LSnorm2}. The extracted values of $M$ were 16588, 16245 and 16697 for EH1, EH2, and EH3, respectively, while $N=16913$. The variations were caused by the different statistics of each hall as modeled in the simulated data sets, and had little impact on the CLs. 

\begin{table}
\begin{center}
\begin{tabular}{cccc}
\hline \hline
  ~Hall~   &~Frequency~(h$^{-1}$)~&~Period~(h)~&~CL~(\%) \\ 
\hline
EH1 & 0.15  & 6.6 & 69.8 \\
EH2 & 0.10  & 10.4 & 5.1 \\
EH3 & 0.11  &  8.9 & 33.9 \\
\hline
\end{tabular}
\caption{Frequency, period and confidence level (CL) of the highest LS power in each hall. The frequency and the period are reported using sidereal hours.}
\label{tab:highestfreqs}
\end{center}
\end{table}

The resulting CL values for each frequency can be seen in the top panels of Fig.~\ref{Fig:LS_CL}. Table~\ref{tab:highestfreqs} gives information about the highest LS power in each EH. It is noteworthy that none of the highest powers is common among the three halls. No significant evidence for a periodic signal was found.

The periodicity search was also performed with the discrete Fourier transform (DFT). Since this method does not account for uneven sampling, the gaps in data acquisition were handled by exploiting the linearity of the transform. The DFT was applied to the data, with the residual survival probability set to zero in all the gap bins (see Fig.~\ref{Fig:Data_hourly}). Many simulated data sets with the same gaps and no time-varying signal were also transformed and averaged for each hall, and then the results were subtracted from the data. The impact of the subtraction was very small, which is expected due to the small number of missed hourly samples (typically, a few per week) relative to the total number of samples (ideally, 168 per week). The resulting power spectra were consistent with those obtained from the LS method.


\section{Analysis on LV-CPTV coefficients }\label{sec:smefit}


The data were also probed for a LV-CPTV signal under the SME. In this framework, the survival probability can be expressed as $P_{\nuebar \to \nuebar} =  P^{(0)} + P^{(1)} + P^{(2)} + ...$. The first term $P^{(0)}  = |S_{\bar{e}\bar{e}}^{(0)}|^2$ is the mass-driven survival probability for $\overline{\nu}_e$'s in the Lorentz-invariant case. $P^{(1)}$ is calculated as~\cite{Neutrino4}
\begin{equation}
\begin{aligned}
P^{(1)}  =&~2L\cdot\Im \biggl[ S_{\idxee}^{(0)*}\sum_{\bar{c},\bar{d}=\bar{e},\bar{\mu},\bar{\tau}}({\cal M}^{(1)}_{\idxee})_{\indx}\cdot \bigl[ \C\indx +\\
& \As\indx\sin\omega_{\oplus}T_{\oplus}+\Ac\indx\cos\omega_{\oplus}T_{\oplus}  + \\
 & \Bs\indx\sin2\omega_{\oplus}T_{\oplus}+\Bc\indx\cos2\omega_{\oplus}T_{\oplus} \bigr] 
 \biggr], 
\label{eq:SMEpred_local}
\end{aligned}
\end{equation}
where $L$ is the baseline, $({\cal M}^{(1)}_{\bar{e}\bar{e}})_{\bar{c}\bar{d}}$ are the so-called experimental factors, $T_{\oplus}$ represents sidereal time and $\omega_{\oplus} = 2\pi$/(1 sidereal day). The subscript $\indx$ runs over the $\idxee$, $\idxmm$, $\idxtt$, $\idxem$, $\idxet$ and $\idxmt$ flavor pairs. $\C\indx$, $\As\indx$, $\Ac\indx$, $\Bs\indx$ and $\Bc\indx$ are commonly referred to as the sidereal amplitudes, which are functions of a total of fourteen SME coefficients for each flavor pair, as well as neutrino energy and propagation direction. The complex relationship between the sidereal amplitudes and the individual coefficients, as well as other details concerning the SME, can be found in the Appendix.  
\begin{figure*}[]
     \includegraphics[ width=0.99\textwidth]{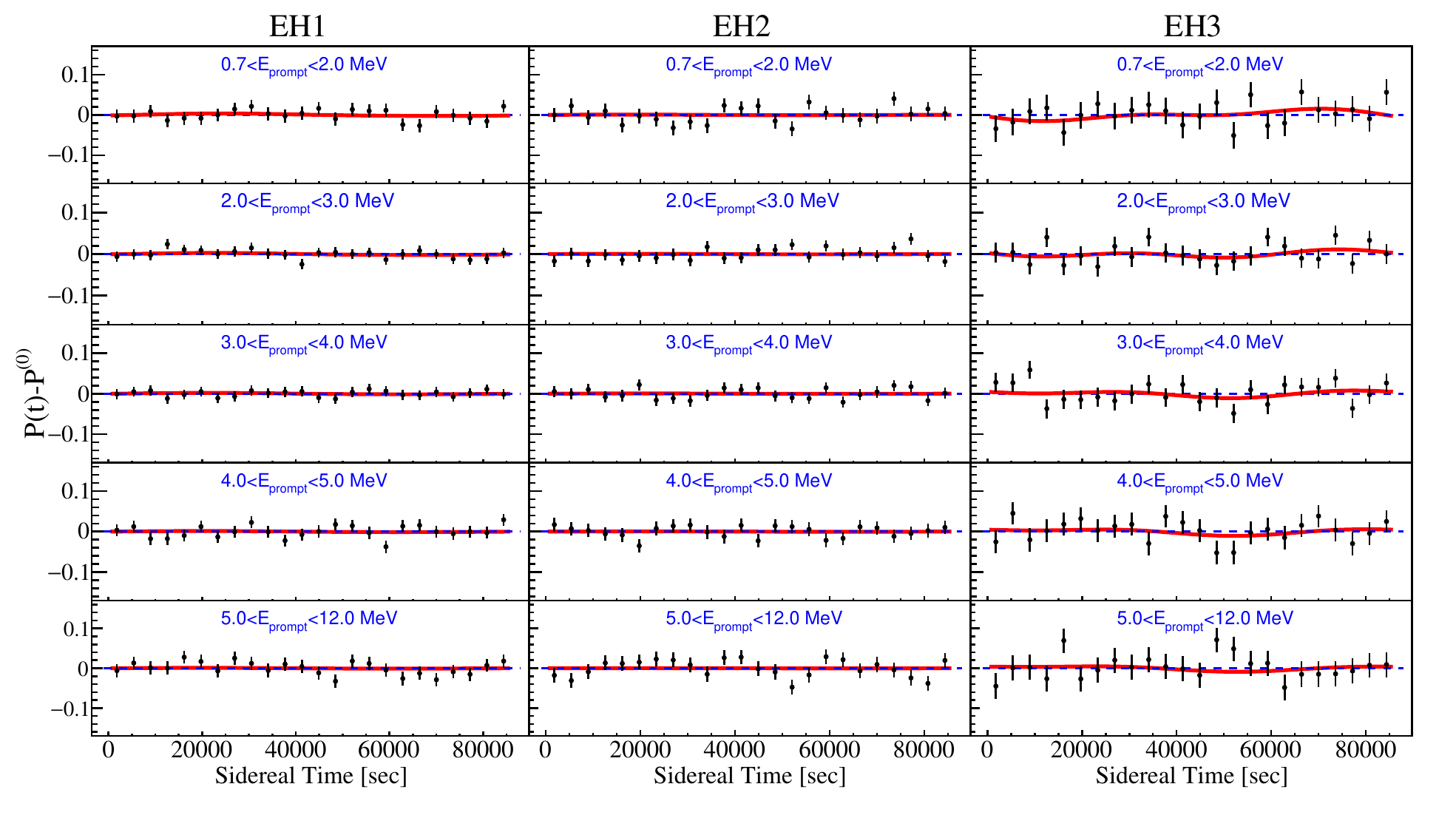}
    \caption{Measured residual survival probability for the three Daya Bay experimental sites and for 5 different prompt energy bins. The best-fit curves for the SME $\idxee$ flavor pair are shown in red.}
\label{Fig:multipanel}
\end{figure*}

The goal of this analysis is to constrain the individual SME coefficients contained in the sidereal amplitudes. For Daya Bay's baselines and energies, $P^{(2)}$ is smaller than $P^{(1)}$ by a few orders of magnitude, and was consequently ignored together with higher order terms. The subtraction of $P^{(0)}$ in $R(t)$ made the fit insensitive to the isotropic amplitude $\C\indx$, whose coefficients can be extracted by analyzing the time-independent energy and baseline dependencies of the oscillation probability. These effects have been constrained by atmospheric neutrino data~\cite{SuperK_Energy,IceCube_Energy} well beyond the reach of Daya Bay. Without $\C\indx$, a total of nine different coefficients are contained in the amplitudes $\As\indx,\Ac\indx,\Bs\indx$ and $\Bc\indx$, as shown in Eq.~(\ref{eq:amp5}). The sum in $P^{(1)}$ over the six flavor pairs makes it unfeasible for a single experiment to simultaneously constrain the $8\times6=48$ parameters with one fit, given their degeneracies. Interplay between the terms could be disentangled by comparing results from experiments with different neutrino energies, directions, and flavors. Without a positive signal however, it is impossible to determine whether there are any correlations or cancellations among the terms within the sum. Accordingly, the standard practice of fitting each flavor pair at a time by setting the coefficients of the other pairs to zero was employed.

Up to now neutrino experiments~\cite{LSNDPRD72_076004,MiniBooNEPLB718,T2K_111101,MINOSPRL101_151601,MINOSPRL105_151601,MINOSPRD85_031101,ICEcubePRD82_112003,DoubleChoozPRD86_112009} have reported limits on the four sidereal amplitudes $\As\indx,\Ac\indx,\Bs\indx$ and $\Bc\indx$. Even when considering individual flavor pairs, the number of direction-dependent parameters precluded these experiments from setting limits on individual coefficients, except through the method of fitting one coefficient at a time while arbitrarily setting all others to zero. With a unique configuration of multiple directions from three experimental sites to three pairs of nuclear reactors (see Fig.~\ref{Fig:Map}) and the separation into five energy bins described below, Daya Bay was able to completely disentangle the energy and direction dependencies in Eq.~(\ref{eq:amp5}) and to simultaneously constrain eight LV-CPTV coefficients: 
$(a_{R})^{X}_{\indx} $, $(c_{R})^{TX}_{\indx}$, $(c_{R})^{XZ}_{\indx}$, $(a_{R})^{Y}_{\indx}$, $(c_{R})^{TY}_{\indx}$,  $(c_{R})^{YZ}_{\indx}$, $(c_{R})^{XX}_{\indx}-(c_{R})^{YY}_{\indx}$ and $(c_{R})^{XY}_{\indx}$. 

The IBD sample was split into five prompt energy bins (0.7, 2.0); (2.0, 3.0); (3.0, 4.0); (4.0, 5.0); and (5.0, 12.0) MeV, chosen so as to contain a similar amount of statistics in each. This resulted in 15 independent data sets (3 EHs $\times$ 5 energy bins), whose residual survival probabilities $R_j(t)=P_j(t)-P_j^{(0)}$ are shown in Fig.~\ref{Fig:multipanel}. Given that each EH sees $\overline{\nu}_e$'s from the six reactor cores, these data sets were simultaneously fit with 
\begin{equation}
R^{\mathrm{fit}}_{j} = \sum_{i} f_{ij} P^{(1)}_{ij}, 
\label{eq:fitfunc}
\end{equation}
where $f_{ij}$ is the expected fraction of events from the $i$th reactor core in data set $j$, and $P^{(1)}_{ij}$ is the oscillation probability of Eq.~(\ref{eq:SMEpred_local}) for that particular $ij$ combination. The event fraction $f_{ij}$ was calculated as
\begin{equation}
f_{ij} = \frac{F_{ij} }{\sum_{k} F_{kj} },
\end{equation}
where $F_{ij}$ is the $i$th core's time-integrated flux seen in the hall corresponding to data set $j$, determined from the reactor power and fission fraction information provided by the power plant~\cite{DYBCPC2017} and including oscillation and inverse-square law effects. Accordingly, the $\chi^2$ used in the fit is expressed as
\begin{equation}
\chi^{2} = \sum_{E_{\mathrm{bin}}= 1}^{5}\sum_{\mathrm{EH}=1}^{3}\sum_{t_{\mathrm{bin}}=1}^{24} \left\{ \frac{(R-R^{\mathrm{fit}})^{2}}{\sigma^{2}_{R}} \right\}_{E_{\mathrm{bin}},\mathrm{EH},t_{\mathrm{{bin}}}}. 
\end{equation}
Here $R=P(t)-P^{(0)}$ is the measured residual survival probability of each data point, $R^{\mathrm{fit}}$ is the SME prediction given by Eq.~(\ref{eq:fitfunc}), and $\sigma^{2}_{R}$ is the total error as described in Sec.~\ref{sec:unc}. The energy spread in each of the five bins was taken into account when calculating $P^{(1)}_{ij}$
and found to be unimportant. Prompt energy was converted to $\overline{\nu}_e$ energy using a response matrix~\cite{DYBPRD2017}. The fit was performed assuming the normal neutrino mass ordering, a zero value for the $CP$-violating phase, and the values of the oscillation parameters reported in Ref.~\cite{PDG2015}. The first two choices, as well as the uncertainties of the oscillation parameters, were found to have a negligible impact on the results. 

\begin{table*}[!htpb] 
\begin{center} 
\caption{Best-fit values $\pm$ 95$\%$ CL limits. NDF stands for the number of degrees of freedom, which corresponds to $3~\mathrm{sites} \times 24~\mathrm{bins} \times 5~\mathrm{energy~bins} - 8~\mathrm{parameters} = 352$. The $\chi^2/$NDF values are very similar because the fit formulas have the same structure, albeit different values of experimental factors $({\cal M}^{(1)}_{\bar{e}\bar{e}})_{\bar{c}\bar{d}}$, resulting in different best-fit parameters and limits. The associated correlation matrices are provided as Supplemental Material~\cite{supmat}.}
\label{tab:BF}
\begin{tabular}{rcccccc} 
\hline 
\hline
Coefficient & $\idxee$ & $\idxmm$ & $\idxtt$ & $\idxem$ & $\idxet$ & $\idxmt$ \\ \hline 
$a^X_R$/$10^{-20}$ (GeV) & $-5 \pm 25$ & $9 \pm 45$ & $13 \pm 58$ & $-3.4 \pm 5.5$ & $-5.6 \pm 8.0$ & $10 \pm 51$ \\
$c^{TX}_R$/$10^{-18}$ & $-15 \pm 55$ & $26 \pm 99$ & $34 \pm 122$ & $-4.5 \pm 7.1$ & $-6.9 \pm 9.7$ & $29 \pm 109$ \\
$c^{XZ}_R$/$10^{-18}$ & $-20 \pm 70$ & $36 \pm 128$ & $43 \pm 153$ & $-2.1 \pm 6.8$ & $-2.7 \pm 8.4$ & $39 \pm 139$ \\
$a^Y_R$/$10^{-20}$ (GeV) & $5 \pm 25$ & $-9 \pm 45$ & $-10 \pm 58$ & $-0.3 \pm 5.5$ & $-0.9 \pm 8.0$ & $-9 \pm 51$ \\
$c^{TY}_R$/$10^{-18}$ & $2 \pm 55$ & $-3 \pm 99$ & $-4 \pm 122$ & $-0.9 \pm 7.1$ & $- 1.6 \pm 9.7$ & $-4 \pm 109$ \\
$c^{YZ}_R$/$10^{-18}$ & $-10 \pm 70$ & $19 \pm 128$ & $22 \pm 152$ & $-1.4 \pm 6.8$ & $-1.9 \pm 8.4$ & $21 \pm 139$ \\
($c^{XX}_R-c^{YY}_R$)/$10^{-18}$ & $13 \pm 46$ & $-24 \pm 84$ & $-29 \pm 103$ & $1.0 \pm 8.2$ & $0.9 \pm 10.5$ & $-26 \pm 92$ \\
$c^{XY}_R$/$10^{-18}$ & $6 \pm 23$ & $-11 \pm 42$ & $-14 \pm 51$ & $1.0 \pm 4.1$ & $1.3 \pm 5.3$ & $-12 \pm 46$ \\ \hline \hline
$\chi^2/$NDF & $318.1/352$ & $318.2/352$ & $318.1/352$ & $315.0/352$ & $313.6/352$ & $318.1/352$ \\ \hline
\end{tabular}
\end{center}
\end{table*}

The two analyses obtained very similar best-fit values and limits, which are shown in Table \ref{tab:BF}. The 95\% CL limits were obtained by constructing an eight-dimensional parameter space and finding the hypervolume enclosing the constant $\chi^{2}$ hypersurface with minimum $\chi^{2}_{\mathrm{\min}}$ plus 15.79 ($\chi^{2} = \chi^{2}_{\mathrm{\min}} + 15.79$). No significant deviations from the Lorentz-conserving scenario were found. Figure~\ref{Fig:multipanel} shows the best-fit curves in the case of the $\bar{e}\bar{e}$ pair as an illustration. Given the higher values of the experimental factors $({\cal M}^{(1)}_{\bar{e}\bar{e}})_{\bar{c}\bar{d}}$ for the $\idxem$ and $\idxet$ flavor pairs in Daya Bay's configuration, the corresponding limits are stronger than for the other cases by about one order of magnitude. These are the first experimental constraints on the coefficients for the $\idxee, \idxmm$ and $\idxtt$ flavor pairs, and are a result of considering the full sum over flavor pairs in $P^{(1)}$ [Eq.~(\ref{eq:SMEpred_local})]. 

\section{Summary}
As a probe of new physics, a model-independent search for a time variation of the reactor $\nuebar$ survival probability was performed with 621 days of Daya Bay data over a period of 704 calendar days. The Lomb-Scargle method yielded no significant evidence for a periodicity in the frequency range of 5.9$\times10^{-5}$ sidereal hour$^{-1}$ to 0.5 sidereal hour$^{-1}$. The survival probability measured at Daya Bay was also examined for a sidereal time dependence within the SME framework. Daya Bay's high statistics and multiple-baseline configuration allowed a complete disentangling of the energy and direction dependencies within the sidereal amplitudes, yielding the first simultaneous constraints of individual Lorentz-violating coefficients for a neutrino experiment. Limits were provided for the $\idxem$, $\idxet$, $\idxmt$, $\idxee$, $\idxmm$ and $\idxtt$ flavor pairs, yielding the first experimental constraints for the latter three. 

\section*{ACKNOWLEDGEMENTS}

We thank J. S. D\'iaz for useful discussions regarding the time-dependent perturbation theory and Lorentz violation in the framework of the SME. Daya Bay is supported in part by the Ministry of Science and Technology of China, the U.S. Department of Energy, the Chinese Academy of Sciences, the CAS Center for Excellence in Particle Physics, the National Natural Science Foundation of China, the Guangdong provincial government, the Shenzhen municipal government, the China General Nuclear Power Group, the Key Laboratory of Particle and Radiation Imaging (Tsinghua University), the Ministry of Education, the Key Laboratory of Particle Physics and Particle Irradiation (Shandong University), the Ministry of Education, the Shanghai Laboratory for Particle Physics and Cosmology, the Research Grants Council of the Hong Kong Special Administrative Region of China, the University Development Fund of the University of Hong Kong, the MOE program for Research of Excellence at National Taiwan University, National Chiao-Tung University, the NSC fund from Taiwan, the U.S. National Science Foundation, the Alfred P. Sloan Foundation, the Ministry of Education, Youth, and Sports of the Czech Republic, the Charles University Research Centre UNCE, the Joint Institute of Nuclear Research in Dubna, Russia, the National Commission of Scientific and Technological Research of Chile, and the Tsinghua University Initiative Scientific Research Program. We acknowledge Yellow River Engineering Consulting Co., Ltd., and China Railway 15th Bureau Group Co., Ltd., for building the underground laboratory. We are grateful for the ongoing cooperation from the China General Nuclear Power Group and China Light and Power Company.

\appendix
\section{Background on the SME}
\label{sec:appendix}

This appendix summarizes the details involved in calculating the SME prediction for the analysis presented in Sec.~\ref{sec:smefit}  and lays out the relationship between the sidereal amplitudes and the individual coefficients. Supplemental Material providing necessary values for the reader to reproduce the results presented in this paper is included online~\cite{supmat}. 

In the SME, the oscillation probability $P_{\nuebar \to \nuebar}$ is given by~\cite{Neutrino4}
\begin{equation}
P_{\nuebar \to \nuebar}  = |S_{\bar{e}\bar{e}}^{(0)} + S_{\bar{e}\bar{e}}^{(1)} + S_{\bar{e}\bar{e}}^{(2)} + ...|^2, 
\end{equation}
where the first three terms of the expansion are 
\begin{equation}
\begin{aligned}
P^{(0)}  &= |S_{\bar{e}\bar{e}}^{(0)}|^2,\\
P^{(1)} &= 2\Re(S_{\bar{e}\bar{e}}^{(0)*}S_{\bar{e}\bar{e}}^{(1)}),\\
P^{(2)} &= 2\Re(S_{\bar{e}\bar{e}}^{(0)*}S_{\bar{e}\bar{e}}^{(2)})+|S_{\bar{e}\bar{e}}^{(1)}|^2.
\label{equ:P0P1P2}
\end{aligned} 
\end{equation}
The transition amplitude $S_{\bar{e}\bar{e}}^{(0)}$ is expressed as 
\begin{equation}
S_{\bar{e}\bar{e}}^{(0)} = \sum_{a}U_{a\bar{e}}^*U^{\phantom{*}}_{a\bar{e}}e^{-iE_{a} L},
\end{equation}
where $U$ is the PMNS~\cite{PMNS} neutrino mixing matrix, $E_{a}$ is the neutrino energy~\cite{supmat}, $L$ is the baseline~\cite{supmat}, and the sum is over all mass eigenstates $a=1,2,3$. $P^{(0)}$ is the usual oscillation probability for massive neutrinos in the Lorentz-invariant case. $P^{(1)}$ and $P^{(2)}$ include the interference from common mass-driven mixing and LV mixing.
$P^{(1)}$ is calculated as~\cite{Neutrino4}
\begin{equation}
\begin{split}
P^{(1)}  = 2L\cdot\Im [S_{\idxee}^{(0)*}\sum_{\bar{c},\bar{d}=\bar{e},\bar{\mu},\bar{\tau}}({\cal M}^{(1)}_{\idxee})_{\indx}\cdot\delta h_{\indx}],
\label{eq:SMEpred}
\end{split}
\end{equation}
where
$({\cal M}^{(1)}_{\bar{e}\bar{e}})_{\bar{c}\bar{d}}$ are the experimental factors~\cite{supmat} and $\delta h_\indx$ is the LV Hamiltonian. The subscript $\indx$ represents the flavor pairs $\idxee$, $\idxmm$, $\idxtt$, $\idxem$, $\idxet$, $\idxmt$. The experimental factors are defined in terms
of the conventional eigenvalues and elements of the PMNS matrix:
\begin{equation}
({\cal M}^{(1)}_{\bar{e}\bar{e}})_{\indx} = \sum_{ab}\tau ^{(1)}_{ab}U^*_{a\bar{e}} U^{\phantom{*}}_{a\bar{c}} U^*_{b\bar{d}} U^{\phantom{*}}_{b\bar{e}},
\end{equation}
where 
\begin{equation}
\tau ^{(1)}_{ab}(E,L) = \left\{
                           \begin{array}{ll}
                              e^{-iE_{b}L}, & \hbox{$E_{a}=E_{b}$}  \\
                              \frac{e^{-iE_{a}L}-e^{-iE_{b}L}}{-i\Delta _{ab}L}, & \hbox{$E_{a}\neq E_{b},$}
                           \end{array}
                         \right.
\label{equ:MeeTauab}
\end{equation}
$b$ runs over all mass eigenstates, and $\Delta _{ab}=E_{a} - E_{b}$ are the standard eigenenergy differences. For Earth-based experiments, the neutrino direction changes with time as both the source(s) and the detector(s) rotate with an angular frequency $\omega_{\oplus} = 2\pi$/(1 sidereal day). The time dependence of the Hamiltonian $\delta h_\indx$ can be parametrized in terms of this sidereal frequency as
\begin{equation}
\begin{aligned}
\delta h_\indx = &\C\indx+\As\indx\sin\omega_{\oplus}T_{\oplus}+\Ac\indx\cos\omega_{\oplus}T_{\oplus}  \\
 +&  \Bs\indx\sin2\omega_{\oplus}T_{\oplus}+\Bc\indx\cos2\omega_{\oplus}T_{\oplus}, 
\end{aligned}
\end{equation}
where $T_{\oplus}$ represents sidereal time. The sidereal amplitudes $\C\indx$, $\As\indx$ and $\Ac\indx$ include both CPTV and LV-CPTV coefficients, while $\Bs\indx$ and $\Bc\indx$ only contain LV-CPTV coefficients~\cite{Neutrino4}, and are determined as
\begin{widetext}
\begin{equation}
\begin{aligned}
\C\indx &= (a_{R})^{T}_{\indx} - \hat{N}^{Z}(a_{R})^{Z}_{\indx} + E\{-\frac{1}{2}(3-\hat{N}^{Z}\hat{N}^{Z})(c_{R})^{TT}_{\indx} +2\hat{N}^{Z}(c_{R})^{TZ}_{\indx} +\frac{1}{2}(1-3\hat{N}^{Z}\hat{N}^{Z})(c_{R})^{ZZ}_{\indx}\}. 
\\
\As\indx &= \hat{N}^{Y}(a_{R})^{X}_{\indx} - \hat{N}^{X}(a_{R})^{Y}_{\indx} +E\{-2\hat{N}^{Y}(c_{R})^{TX}_{\indx} +2\hat{N}^{X}(c_{R})^{TY}_{\indx}  +2\hat{N}^{Y}\hat{N}^{Z}(c_{R})^{XZ}_{\indx}  -2\hat{N}^{X}\hat{N}^{Z}(c_{R})^{YZ}_{\indx}\},
\\
\Ac\indx &= -\hat{N}^{X}(a_{R})^{X}_{\indx} - \hat{N}^{Y}(a_{R})^{Y}_{\indx}  +E\{2\hat{N}^{X}(c_{R})^{TX}_{\indx}+2\hat{N}^{Y}(c_{R})^{TY}_{\indx}  -2\hat{N}^{X}\hat{N}^{Z}(c_{R})^{XZ}_{\indx}  -2\hat{N}^{Y}\hat{N}^{Z}(c_{R})^{YZ}_{\indx}\},
\\
\Bs\indx &= E\{ \hat{N}^{X}\hat{N}^{Y}((c_{R})^{XX}_{\indx}-(c_{R})^{YY}_{\indx})-(\hat{N}^{X}\hat{N}^{X} -\hat{N}^{Y}\hat{N}^{Y})(c_{R})^{XY}_{\indx}\},
\\
\Bc\indx &= E\{ -\frac{1}{2}(\hat{N}^{X}\hat{N}^{X} -\hat{N}^{Y} \hat{N}^{Y} )((c_{R})^{XX}_{\indx}-(c_{R})^{YY}_{\indx})-2\hat{N}^{X}\hat{N}^{Y} (c_{R})^{XY}_{\indx}\}. 
\label{eq:amp5}
\end{aligned}
\end{equation}
\end{widetext}
Here $(T,X,Y,Z)$ denote the coordinates of the sun-centered celestial-equatorial reference frame, and $\hat{N}^{X}, \hat{N}^{Y}, \hat{N}^{Z}$ are the directional factors, defined as
\begin{equation}
\begin{pmatrix}
\hat{N}^{X}\\
\hat{N}^{Y}\\
\hat{N}^{Z}\\
\end{pmatrix}
=
\begin{pmatrix}
\cos\chi \sin\theta\  \cos\phi + \sin\chi \ \cos\theta \\
\sin\theta \ \sin\phi \\
-\sin\chi \ \sin\theta\  \cos\phi +\cos\chi\  \cos\theta
\end{pmatrix}, 
\label{eq:df}
\end{equation}
where $\chi$ is the laboratory colatitude (the polar angle measured from the north), $\theta$ is the angle between the neutrino beam and the local zenith, and $\phi$ is the angle between the beam and east of south~\cite{supmat}. A total of 14 SME LV coefficients are contained in Eq.~(\ref{eq:amp5}) for each flavor pair: $(a_{R})^{T}_{\indx}$,  $(a_{R})^{Z}_{\indx}$, $(c_{R})^{TT}_{\indx}$, $(c_{R})^{TZ}_{\indx}$, $(c_{R})^{ZZ}_{\indx}$, $(a_{R})^{X}_{\indx} $, $(c_{R})^{TX}_{\indx}$, $(c_{R})^{XZ}_{\indx}$, $(a_{R})^{Y}_{\indx}$, $(c_{R})^{TY}_{\indx}$,  $(c_{R})^{YZ}_{\indx}$, $(c_{R})^{XX}_{\indx}$, $(c_{R})^{YY}_{\indx}$ and $(c_{R})^{XY}_{\indx}$. The nine coefficients included in the $\As\indx,\Ac\indx,\Bs\indx$ and $\Bc\indx$ amplitudes are the ones constrained in the analysis of Sec.~\ref{sec:smefit}. It should be noted that the coefficients for left-handed neutrinos are related to those for right-handed antineutrinos via 
$(a_{R})^{\alpha}_{\bar{c}\bar{d}} = -(a_{L})^{\alpha*} _{cd}$ and $(c_{R})^{\alpha \beta}_{\bar{c}\bar{d}} = (c_{L})^{\alpha\beta *} _{cd}$, with $\alpha,\beta = T,\ X,\ Y,\ Z$~\cite{Neutrino4}.

\end{document}